\documentstyle[12pt,epsfig]{elsart}
\def\Journal#1#2#3#4{{#1} {\bf #2}, #3 (#4)}

\def\NIMA{{\em Nucl. Instrum. Methods} A}
\def\NIMB{{\em Nucl. Instrum. Methods} B}

\def\NPBP{{\em Nucl. Phys.} B (Proc. Suppl.)}
\def\NPA{{\em Nucl. Phys.} A}
\def\PLB{{\em Phys. Lett.}  {\bf B}}
\def\PRL{\em Phys. Rev. Lett.}
\def\PRD{{\em Phys. Rev.} {\bf D}}
\def\PRC{{\em Phys. Rev.} C}

\def\PRP{{\em Phys. Rep. }}

\def\ZPC{{\em Z. Phys.} C}
\def\ZPA{{\em Z. Phys.} A}

\def\PAN{\em Phys. Atom. Nucl.}

\def\PNPP{\em Prog. Nucl. Part. Phys.}

\def\PTPS{\em Prog. Theo. Phys. Suppl.}

\def\APP{\em Astroparticle Physics}
\def\EPL{\em Europhys. Lett.}

\def\JPG{\em Journal of Physics G}

\def\ANDT{\em Atom. Nucl. Dat. Tab.}

\def\ra{\rightarrow}
\def\be{\begin{equation}}
\def\ee{\end{equation}}

\newcommand{\ls}{\mbox{$\stackrel{<}{\sim}$ }}
\newcommand{\expe}{experiment}

\newcommand{\osz}{oscillations}

\newcommand{\bb}{double beta decay}
\newcommand{\obb}{0\mbox{$\nu\beta\beta$-decay}}
\newcommand{\zbb}{2\mbox{$\nu\beta\beta$-decay}}
\newcommand{\nbb}{neutrinoless double beta decay}
\newcommand{\majo}{Majorana}

\newcommand{\bmbm}{\mbox{$\beta^-\beta^-$} }
\newcommand{\bpbp}{\mbox{$\beta^+\beta^+$} }
\newcommand{\ecec}{\mbox{$EC/EC$} }
\newcommand{\bec}{\mbox{$\beta^+/EC$} }

\newcommand{\bnel}{\mbox{$\bar{\nu}_e$} }

\newcommand{\nel}{\mbox{$\nu_e$}}

\newcommand{\neu}{neutrino}
\newcommand{\neus}{neutrinos}

\newcommand{\ema}{\mbox{$\langle m_{ee} \rangle$ }}

\newcommand{\gess}{\mbox{$^{76}Ge$ }}

\newcommand{\zns}{\mbox{$^{70}Zn$ }}

\newcommand{\cd}{\mbox{$^{116}Cd$ }} 
\newcommand{\cdhdz}{\mbox{$^{113}Cd$ }}
\newcommand{\cdhs}{\mbox{$^{106}Cd$ }} 

\newcommand{\cdhvz}{\mbox{$^{114}Cd$ }} 

\newcommand{\tehzw}{\mbox{$^{120}Te$ }} 
\newcommand{\teha}{\mbox{$^{128}Te$ }} 
\newcommand{\tehd}{\mbox{$^{130}Te$ }}
\newcommand{\tehdz}{\mbox{$^{123}Te$ }}

\newcommand{\ba}{\begin{array}{c}}
\newcommand{\baz}{\begin{array}{cc}}
\newcommand{\bad}{\begin{array}{ccc}}
\newcommand{\bea}{\begin{equation} \begin{array}{c}}
\newcommand{\eea}{ \end{array} \end{equation}}
\newcommand{\ea}{\end{array}}

\begin{document}
\begin{frontmatter}
\title{COBRA - Double beta decay searches using CdTe detectors}
\author{K. Zuber}
\address{Lehrstuhl f\"ur Experimentelle Physik IV, Universit\"at Dortmund,
Otto-Hahn Str.4, 44221 Dortmund, Germany}
\begin{abstract}
A new approach (called COBRA) for investigating \bb{} using CdTe semiconductor
detectors is proposed. This will allow simultaneous measurements
of 5 $\beta^-\beta^-$ - and 4 $\beta^+\beta^+$ - emitters at once.
Half-life limits for neutrinoless double beta decay of \cd and \tehd{} 
can be improved by more than one order
of magnitude and sensitivities on the effective Majorana neutrino mass of less than
1 eV can be obtained. 
Furthermore, for the first time a realistic chance of observing 
double electron capture processes exists. Additional searches for rare processes
like the 4-fold forbidden \cdhdz{} $\beta$-decay, the electron capture of \tehdz
and dark matter detection can be performed. 
The achievable limits are evaluated for 10 kg of such detectors and can be
scaled accordingly towards higher masses because of the modular design of the proposed experiment.
\end{abstract}
{\small PACS: 13.15,13.20Eb,14.60.Pq,14.60.St}
\begin{keyword}
massive neutrinos, double beta decay
\end{keyword}
\end{frontmatter}

\section{Introduction}
The fundamental question whether \neus{} have a non-vanishing rest mass
is still one of the big open problems of particle physics.
In case of massive neutrinos a variety of new physical processes open up
\cite{zub98}.
Over the last years evidence has grown for a non-vanishing mass by investigating
solar and atmospheric neutrinos as well as by results coming from
the LSND-\expe{}.
They all can be explained within the framework of \neu{} \osz{}
\cite{bil99}.
However oscillations only depend on the differences of squared masses and are therefore
no absolute mass measurements.
Besides, the question concerning the fundamental character of \neus{}, whether
being Dirac- or Majorana particles, is still unsolved.
A process contributing information to both questions is \nbb{} of a nucleus
($Z,A$) 
\be
(Z,A) \ra (Z+2,A) + 2 e^-  \quad (\obb)
\ee
This process is violating lepton-number by two units and only allowed if
\neus{} are massive \majo{} particles.
The quantity which can be extracted out of this is called effective \majo{} \neu{} mass
\ema{} and given by
\be
\label{eq:ema}\ema = \mid \sum_i U_{ei}^2 \eta_i m_i \mid
\ee
with the relative CP-phases $\eta_i = \pm 1$, $U_{ei}$ as the mixing
matrix elements and
$m_i$ as the
corresponding mass eigenvalues.
Currently the best limit is given by investigating \gess{} resulting in an
upper bound of \ema $\ls$ 0.35 eV \cite{bau99}. 
Moreover the standard model process
\be
(Z,A) \ra (Z+2,A) + 2 e^- + 2 \bnel \quad (\zbb)
\ee
can be investigated as well. It is important to check the
reliability of the calculated
nuclear matrix elements.\\
The experiment proposed here utilizes Cd-based semiconductor detectors.
A big benefit of this approach is, that the double beta emitters are
part of the detectors itself.
The detectors can be used either in the form of only measuring the sum energy
of both electrons or in a modified way as pixelised detectors, which allows
simultaneous tracking and energy
measurement. CdTe and CdZnTe (CZT) detectors are
used in several areas
of X-ray physics, astrophysics and medical applications. 
All isotopes of interest for \bmbm{}-decay searches which
are intrinsic in these detectors are listed in Tab.1.
A few observations exist for \zbb{} of \cd, most of them with very low statistics.
The obtained half-lives center around $3 \cdot 10^{19}$ a \cite{kum94,arn96,dan00}. 
The \zbb{} of \tehd
was only measured by geochemical methods and found to be in a range of $0.7 - 2.7 \cdot 10^{21}$ a, 
therewith a spread
of a factor four
among different measurements exist \cite{kir86,man91,ber92,tak96}. 
Current lower limits on \obb{} half-lives for the two favourite isotopes \cd and \tehd are
$7 \cdot 10^{22}$ a and $1.44 \cdot 10^{23}$ a (both 90 \% CL) respectively
\cite{dan00,ale00}. Limits on the other mentioned isotopes are rather poor, a
compilation can be found in \cite{tre95}.\\ 
In a more general scheme \obb{} can be realised by
several lepton number violating mechanisms. Beside massive Majorana \neu{} exchange,
additional mechanisms like
right-handed weak currents, R-parity violating supersymmetry, double charged
Higgs bosons or leptoquarks have been proposed \cite{moh81,hir95,hir96}.\\
To obtain more information on the underlying physics process, it is worthwile
to look for
transitions into excited states \cite{doi85} and to
investigate \bpbp{}-decay \cite{hir94}.
Three different decay channels can be considered here
\bea
(Z,A) \ra (Z-2,A) + 2 e^+ + (2 \nel) \quad \mbox{(\bpbp{})}\\
e- + (Z,A) \ra (Z-2,A) + e^+ + (2 \nel) \quad \mbox{(\bec{})}\\ 
2 e^- + (Z,A) \ra (Z-2,A) + (2 \nel) \quad \mbox{(\ecec{})}
\eea
where the last two cases involve electron capture.
The first two are the easiest to detect because of the annihilation photons of the positron(s) emitted. 
On the other hand they are largely suppressed by phase space reduction ($Q-4m_e c^2$ for \bpbp{} and $Q-2
m_ec^2$ 
for \bec{} respectively).
The one with the lowest expected half-life is \ecec{}- decay, but difficult to
detect
because only X-rays are emitted. All \bpbp{}-decays can occur as \zbb{} or \obb{}.
Current half-life limits for the decay modes involving at least one
positron are of the order of 10$^{20}$ a 
\cite{bel99,gav00}, far below theoretical expectation. 
The proposed experiment would for the first time allow the measurement of \ecec{}-decay
in the theoretically expected range
because the isotopes are within the CdTe crystals (source equal to detector) 
and do not rely on external devices for
detecting 511 keV photons.
All isotopes of interest for such studies are given in Tab. 2.
Moreover, transitions to excited final states for most of the above listed isotopes are possible.
This includes basically $\gamma$-rays of 1.294 MeV and 1.757 MeV (\cd), 512 keV, 1.13 MeV and
1.56
MeV (\cdhs) for Cd-isotopes and 442.9
keV (\teha), 536.1 keV (\tehd) and 1.171 MeV (\tehzw) for tellurium. For a compilation of existing
limits
see \cite{tre95}.
A first small attempt to use CdTe for rare decay searches was done by \cite{mit88}.

\section{Experimental considerations}
The proposed experiment, running under the name COBRA\footnote{CdTe 0 neutrino double Beta
Research Apparatus},
should consist in a first stage
of 10 kg of material in form of either CdTe or CdZnTe detectors.
For double
beta decay searches especially two experimental parameters have to be primarily considered, namely
energy resolution and the 
expected background. A possible setup is shown schematically in Fig.1.
The central detector is an array of about 1600 CdTe crystals, because the largest
available crystal size is of the order of 1 cm$^3$.
In case of a cubic arrangement it would have a size of 12$\times$12$\times$12 cm$^3$.
The detectors will be installed within a NaI detector for two reasons:
First, it can act as an active veto against penetrating particles and 
secondly it can be used as a detector for $\gamma$-rays, coincidence measurements
between a CdTe detector and the NaI can be done.
It has been demonstrated 
by several dark matter groups 
that such  detectors can be built as low-level devices 
\cite{dama,ukdm}. Coincidences among the various CdTe crystals can be formed as well.\\ 
This inner part is surrounded by a shield of very clean oxygen free high conductivity (OFHC) copper in
combination with
low-level lead,
a setup common to low-level experiments. The complete apparatus
will be covered by an active veto against muons made out of high efficiency 
scintillators. Clearly the experiment should be located in one of the existing underground
facilities.
A further shielding against thermal neutrons might be necessary, because
of the large cross section of $^{113} Cd (n,\gamma) ^{114} Cd$ reactions.\\
Common background 
to all low level experiments are
the uranium and thorium decay chains as well as $^{40}K$ contaminations. Some parts of
the chains might be eliminated
by subsequent detections within one crystal, clearly identifying the origin of the event.
As an example take a sequence from the $^{226}$Ra decay chain (from the $^{238}$U decay):\
\begin{eqnarray*}
& & ^{214}Bi (Q_\beta = 3.27 MeV, T_{1/2} = 19.9 \mbox{min}) \ra ^{214}\mbox{Po}\\
& & ^{214}\mbox{Po} (Q_\alpha = 7.83 MeV, T_{1/2} = 164.3 \mu s)  \ra ^{210}Pb
\end{eqnarray*}
This $\beta - \alpha$ coincidence within
one crystal can be used to estimate this background contribution.
Studies on radioisotope production in CdTe due to
cosmic ray activation have been performed using proton beams
\cite{por96}.\\
The intrinsic background from \cdhdz{}-decay is of no concern
because its endpoint is around 320 keV, much below the \obb{} lines. 
A measurement of the energy spectrum in the range 2-3 MeV obtained with a test setup
using a conventional 1 cm$^3$ CdTe
is shown in Fig.2.\\
The smallness of the CdTe detectors makes it possible
to construct the experiment in a modular design, making future upgrades easy.\\
The principle readout of one CdTe detector will focus on electron collection only, to avoid
smearing in the energy because of the bad hole mobility in CdTe.
Energy resolutions of about 1 \% for the $^{137}$Cs line at 662 keV 
have already been achieved \cite{he99}. This is sufficient to assure no overlap between
the $0 \nu$ region of \tehd with possible background lines at 2447.7 keV ($^{214}$Bi) and 2614.4 keV
($^{208}$Tl).
A further improvement of the energy resolution might be achieved by a slight cooling of
the CdTe detectors to temperatures of roughly $-$ 20 degrees centigrade.\\ 
As a modification the usage of pixelized detectors is envisaged.
In addition to an energy measurement 
this allows tracking of the two emitted electrons and therefore a handle on background reduction.
However, it has to be investigated experimentally 
in more detail whether the additional pixel bonds are not causing
relatively more impurities. On the other hand this is a very attractive method to search
for transitions to final states using separate pixels within one crystal. Pulse shape analysis techniques
might be performed as well for background subtraction. 
A more detailed treatment of experimental details and simulation studies can be
found in \cite{mum}.

\section{Expected sensitivities for \bmbm{}- decays}
The dominant $2 \nu \beta \beta$-decays are coming from \cd and \tehd. 
With an assumed half-life
of $3 \cdot 10^{19}$ a this produces a count rate of 94 events/day.
Therefore a high statistics measurement of this decay is possible.
For \tehd with an assumed half-life range of $0.7 - 2.7 \cdot 10^{21}$ a 6 - 23 events/day
are expected.
This corresponds to a 6.4 \% - 24.5 \% contribution to the \cd spectrum. Clearly a decision
among the lower and higher half-lives for \tehd can be made.
A detection and proof of the geochemical half-life obtained for \teha will be very difficult,
because it would roughly result in only about 1 event/a. The rather poor limit of
$T_{1/2}^{2\nu} > 9.2 \cdot 10^{16}$ a for \cdhvz{} can certainly be improved, in case of using CZT
for the first
time a limit on the \zbb{} of \zns can be obtained.\\
With regard to neutrino mass limits again the main focus lies on \cd and \tehd.
Achievable half-life limits after 5 years of measurement are shown in Fig.3.
In case of building a background free detector, the corresponding half-lives 
scale linear with measuring time, resulting in an even better sensitivity.
Using the parameters given in Tab.1, a limit on \ema in the region
below 1 eV can be achieved. For an extensive discussion of the
status of the necessary nuclear matrix element calculations see \cite{suh98,fae98}.
If no signal is observed, the observed limit on \ema  would strengthen the believe in
the result
already obtained with Ge, because of the uncertainties coming from nuclear matrix 
element calculations. 
Clearly a further improvement can be achieved by adding more detectors, which
is possible because of the modular design of the experiment.\\
Also transitions to excited states can be investigated because of the good
sensitivities of CdTe detectors for $\gamma$-rays.
Current bounds for such transitions are in the order of $10^{21}$ a \cite{bel87,pie94}.
The modular layout of this experiment would allow to perform a high sensitivity
search. The coincident detection of the deexcitation photon in one CdTe crystal and
the corresponding electron signal in a neighbouring detector forms a clear signal.
This will significantly reduce the background in searches
for these channels.

\section{Experimental sensitivities for \bpbp{}-decays}

A wide range of results can be obtained for the various decay modes of 
\bpbp-decays given in Tab. 2. 
As already stated, the
lowest expected half-life belongs to  the \ecec - decay mode. 
The filling up of the two K-shell holes in $^{106,108}$Pd coming from the decay of
Cd-isotopes will result in
a peak at 48.6 keV. 
A recent calculation for \cdhs \ecec results in a theoretical predicted half-life of $4 \cdot 10^{20}$ a
\cite{suh01}. The expected count rate then is about 550 events/a,
which should result in a clear observation. 
A corresponding peak for the \ecec of $^{120}$Te to $^{120}$Sn would be at 58.4 keV.
There is only a limit on the \bec{} of $4.2 \cdot 10^{12}$ a \cite{fre52},
which can be improved by many orders of magnitude, because the expected number for COBRA is $1.2 \cdot
10^{7}$ events/day.
Transitions to excited states for \cdhs and $^{120}$Te can be explored in a similar fashion.
While limits of the order of $10^{18}$ a exist for \cdhs \cite{dan95}, nothing is known so far
for \tehzw. 

\section{Additional physics - dark matter searches and \cdhdz, \tehdz -decay}
As most low-level double beta detectors also CdTe could be used for dark matter
searches. Detectors with thresholds of about 1 keV at a temperature of $-$ 20 degrees are available. From
the
theoretical point of view, $^{125}$Te together with $^{129}$Xe is among the theoretically most preferred
isotopes to study spin-dependent interactions \cite{res97,tov00}. With 10 kg of CdTe it will be possible
to probe 
the DAMA evidence \cite{bel00} within reasonable time scales. 
Unfortunately no theoretical calculation for the usage of Cd-isotopes for dark matter searches
exists.\\
A long standing discussion is
connected with the $\beta$-decay of \tehdz{}.
This second forbidden unique electron capture occurs with a transition energy of 51.3 $\pm$ 0.2
keV to
the ground state
of $^{123}$Sb.
Measurements concentrating on the detection of the 26.1 keV photons of
K X-rays from $^{123}$Sb
resulted in a 
half-life of the order 10$^{13}$ years \cite{wat62}. However, a new measurement claims a
value of
2.4 $\pm$ 0.9 $\cdot 10^{19}$ a \cite{ale96}, six orders of magnitude higher. The discrepancy might be
associated with
confusing
the above X-ray line
with the Te K X-ray line at 27.3 keV.
Having a decay within the CdTe detector itself, in contrast to the above measurements, this problem can be
solved because
the full 
transition energy can be measured with high efficiency and good energy resolution. Even the long half-life
would correspond to 19 decays per day.\\
Last but not least there is the $\beta$-decay of \cdhdz{}.
This 4-fold
forbidden decay has an uncertain half-life of about $8 \cdot 10^{15}$ a and a Q-value
of about 320 keV. Two measurements exist \cite{ale94,dan95}, but are within their errors in slight
disagreement.
In the COBRA-setup discussed here a very high statistics measurement can be done, having about
10 decays per second with good energy resolution.

\section{Summary and conclusion}
In this paper the physics potential of CdTe or
CdZnTe detectors for double beta decay searches and other rare processes is discussed. 
CdTe detectors profit from the fact
that source material and detector are identical.
An experimental advantage of
the described setup is the good energy resolution (in contrast to scintillators) also in combination with
possible tracking and the possibility to perform
the experiment at room temperature or only slightly below (in contrast to cryogenic detectors).
The unique chance
of investigating in total 5(4) $\beta^-\beta^-$ and 4(3) $\beta^+\beta^+$ - emitters
at the same time in case of using CZT (CdTe)
can be realised.
For neutrinoless double
beta decay an improvement
on the existing half-life limits for the most promising isotopes 
\cd and \tehd by more
than one order of magnitude with respect to current limits could be obtained. 
Thus for both
isotopes a neutrino mass limit of \ema \ls
1 eV would result. A high statistics measurement of the $2\nu \beta \beta$-decays
of both isotopes is possible. Furthermore, a detailed investigation
of \bpbp-decays can be done and for the first time an attempt to measure
\ecec-decays in the theoretically predicted range can be performed.
Sensitive searches for a large number of excited state transitions are feasible as well.
As further topics a high statistics measurement of the 4-fold forbidden
$\beta$-decay of \cdhdz can be conducted, the six orders of magnitude discrepancy for
the electron capture of \tehdz can be solved and a sensitive search for
dark matter can be done. The given numbers scale with the used mass
and because of the modular design of the experiment a corresponding
upgrade for improvements is possible as discussed for a 100 kg solution.

\section{Acknowledgements}
I would like to thank Y. Ramachers
for valueable discussions, D. M\"unstermann for working together on building up the
test setup and performing the
measurements and C. G\"o{\ss}ling for his support. Also I thank
the mechanical workshop and T. Villett of the University
of Dortmund for their help in building the test setup.

\newpage
\begin{center}
\begin{table} [hhh]
\begin{tabular}{|c|c|c|c|c|c|c|}
\hline
Isotope & Q(keV) & nat. ab. (\%) & G$_{2\nu}^{-1}$ (a)& G$_{0\nu}^{-1}$ (a) & ${\bar T}_{2\nu}
(a)$ & $T_{1/2}
\cdot m_{\nu} (a \cdot eV^2)$ \\
\hline
$^{70}$Zn & 1001 & 0.62 & 3.17E21& 4.27E26 & 3.99E22 & 9.83E25\\
$^{114}$Cd & 534  & 28.7 & 6.93E22 & 6.1E26 & 1.74E24 & 5.07E25 \\
$^{116}$Cd & 2805 & 7.5 & 1.25E17 & 5.3E24& 6.31E19 & 4.87E23\\
$^{128}$Te & 868 & 31.7 & 1.18E21 & 1.46E26 & 2.63E24 & 7.77E24\\
$^{130}$Te & 2529 & 33.8 & 2.08E17 & 5.89E24 & 1.84E21 & 4.89E23\\
\hline
\end{tabular}
\bigskip\\
\caption{$\beta^-\beta^-$-isotopes of relevance for \bb{} in CdTe and CZT
detectors.
Given are the Q-values of the transition, the natural abundance, phase space factors and
theoretical
expectations for the half-lives. The \zbb{} half-life is averaged over an acceptable range
of the adjustable parameter $g_{pp}$. The theoretical predictions are taken from \protect
\cite{ast}.}
\end{table}
\end{center}
\begin{center}
\begin{table}[hhh]
\begin{tabular}{|c|c|c|c|}
\hline
Isotope & Q(keV) & nat. ab. (\%) & Decay modes\\
\hline
$^{64}$Zn & 1096.3 & 48.6 & \ecec{}, \bec \\
$^{106}$Cd & 2771 & 1.25 & \ecec{}, \bec{}, \bpbp \\
$^{108}$Cd & 231 & 0.9 &  \ecec \\
$^{120}$Te & 1722 & 0.10 & \ecec{}, \bec \\
\hline
\end{tabular}
\bigskip\\
\caption{$\beta^+\beta^+$ - isotopes of relevance in CdTe and CZT
detectors.
Given are the Q-values of the transition and the natural abundance.
Also given are the possible decay modes.}
\end{table}
\end{center}

\newpage
\begin{center}
\begin{figure}
\epsfig{file=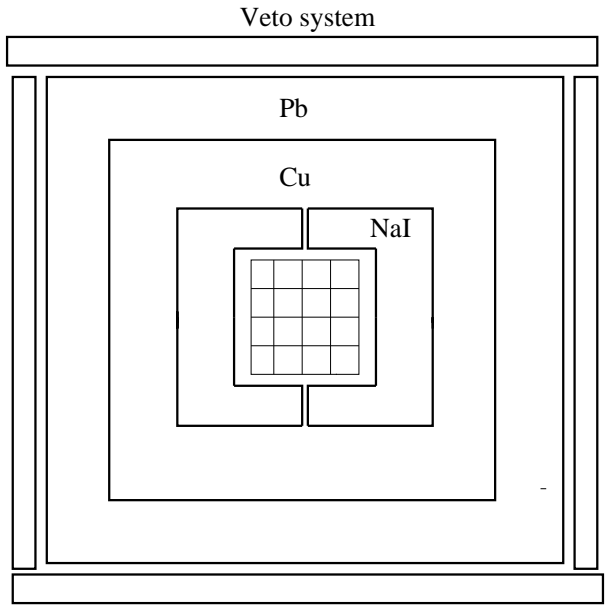,height=8cm,width=8cm}
\vspace{2mm} 
\caption{Schematic layout of the proposed COBRA experiment. An array of CdTe detectors is installed
within two NaI detectors, serving as active veto and part of coincidence measurements.
This will be installed inside an OFHC copper shield, surrounded by low level lead. 
As a veto the complete setup will be surrounded by a muon veto consisting of high efficiency 
plastic scintillators.}
\end{figure}
\end{center}
\begin{center}
\begin{figure} 
\epsfig{file=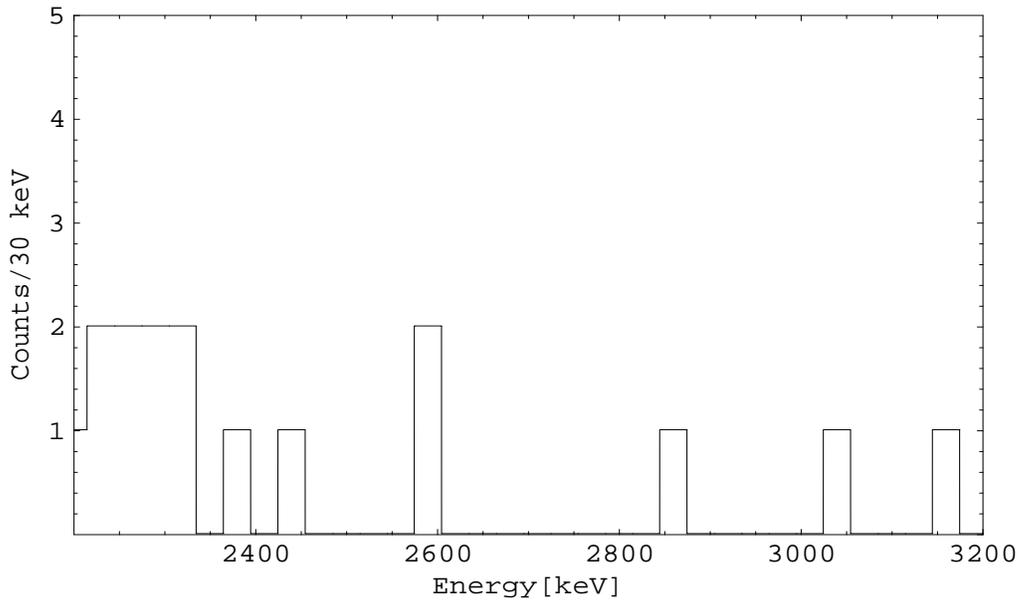,height=8cm,width=14cm}
\vspace{2mm}
\caption{Energy spectrum of a 1 cm$^3$ CdTe detector in the interesting range of 2.2 - 3.2 MeV.
The measuring time corresponds to 135.5 hours. The detector was installed in a shielding of 
10 cm standard grade copper surrounded by an additional shield of 20 cm of spectroscopy lead.
The whole apparatus was surrounded by a $4\pi$ veto made of plastic scintillators.
The total shielding depth was about 5 mwe. For more details see \protect \cite{mum}.
Expected \obb lines are at 2529 keV (\tehd) and 2805 keV (\cd).}
\end{figure}
\end{center}
\begin{center}
\begin{figure} 
\epsfig{file=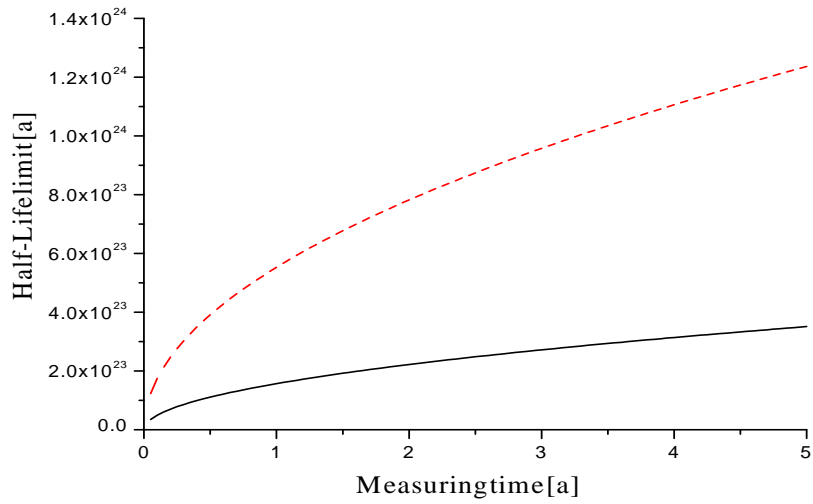,height=8cm,width=12cm}
\vspace{2mm}
\caption{Expected half-life limits for \cd (solid line) and \tehd{}(dashed line) as a function of measuring
time.
Assumed are the experimentally obtained background levels of 0.2 counts/keV/kg/a (\tehd{} 
\protect \cite{ale00}) and
0.03 counts/keV/kg/a (\cd \protect \cite{dan00}) as well as an energy resolution of 1 \%.}
\end{figure}
\end{center}

\end{document}